\documentclass{aastex63}

\shorttitle{Radiation Belt Response to Fast Reverse Shock at Geosynchronous Orbit}
\shortauthors{Bhaskar et al.}
\graphicspath{{./}{figures/}}

\begin{document}

\title{Radiation Belt Response to Fast Reverse Shock at Geosynchronous Orbit}

\correspondingauthor{Ankush Bhaskar}
\email{ankushbhaskar@gmail.com}

\author[0000-0003-4281-1744]{Ankush Bhaskar}
\affiliation{NASA/Goddard Space Flight Center, Greenbelt, Maryland, United States}
\affiliation{University Coperation of Atmospheric Research, Boulder, United States}
\affiliation{Space Physics Laboratory, Vikram Sarabhai Space Centre, Thiruvananthapuram, India}

\author{David Sibeck}
\affiliation{NASA/Goddard Space Flight Center, Greenbelt, Maryland, United States}

\author{Shrikanth G. Kanekal}
\affiliation{NASA/Goddard Space Flight Center, Greenbelt, Maryland, United States}

\author{Howard J. Singer}
\affiliation{NOAA-Space Weather Prediction Center,Colorado, United States}

\author{ Geoffrey Reeves}
\affiliation{Los Alamos National Laboratory,New Mexico, United States}

\author{ Denny M. Oliveira}
\affiliation{NASA/Goddard Space Flight Center, Greenbelt, Maryland, United States}
\affiliation{Goddard Planetary Heliophysics Institute, University of Maryland, Baltimore, United States}

\author{ Suk-Bin Kang}
\affiliation{NASA/Goddard Space Flight Center, Greenbelt, Maryland, United States}

\author{ Colin Komar}
\affiliation{NASA/Goddard Space Flight Center, Greenbelt, Maryland, United States}



\begin{abstract}
	Fast reverse shocks (FRSs) cause the magnetosphere to expand, by contrast to the  well-known compressions caused by the impact of fast forward shocks (FFS). Usually, FFSs are more geoeffective than FRSs, and consequently the inner magnetosphere dynamic responses to both shock types can be quite different. In this study, we investigate for the first time the radiation belt response to an FRS impact using multi-satellite observations and numerical simulations. Spacecraft on the dayside observed decreases in magnetic field strength and energetic ($\sim$40-475 keV) particle fluxes. Timing analysis shows that the magnetic field signature propagated from the dayside to the nightside magnetosphere. Particles with different energies vary simultaneously at each spacecraft, implying a non-dispersive particle response to the shock. Spacecraft located at lower L-shells did not record any significant signatures. The observations indicate a local time dependence of the response associated with the shock inclination, with the clearest signatures being observed in the dusk-midnight sector. Simulations underestimate the amplitude of the magnetic field variations observed on the nightside. The observed decreases in the electron intensities result from a combination of radial gradient and adiabatic effects. The radial gradients in the spectral index appear to be the dominant contributor to the observed variations of electrons seen on the dayside (near noon and dusk) and on the nightside (near midnight). This study shows that even an FRS can affect the radiation belts significantly and provides an opportunity to understand their dynamic response to a sudden expansion of the magnetosphere.
	
\end{abstract}



\section{Introduction} \label{sec:intro}

Interplanetary (IP) shocks are a recurrent feature of the solar wind \citep{Burlaga1971a,Richter1985}. Fast mode IP shocks occur when the relative speed between the ambient solar wind and the shock speed is larger than the local magnetosonic speed \citep{Landau1960,Richter1985,Tsurutani2011a,Oliveira2017a}. Fast IP shocks that are move away from the Sun are classified as fast forward shocks (FFSs), while fast shocks that propagate towards the Sun are classified as fast reverse shocks (FRSs) \citep{Burlaga1995,Tsurutani2011a,Oliveira2017a}. Similarly to FFSs, FRSs are carried antisunward by the continuous solar wind  flow as seen from a reference frame defined as a spacecraft in the IP space or the Earth itself \citep{Richter1985,Burlaga1995,Tsurutani2011a,Oliveira2017a,Oliveira2018a}. FFSs are more numerous during solar maxima, while FRS occurrence rates have no clear correlation with solar activity\citep{Echer2003a,Kilpua2015,Cavus2019}. \cite{Kilpua2015} showed that the occurrence rates of FFSs are higher than the occurrence rates of FRSs during all solar phases, except during solar minima. \par

The steepening conditions across the fronts of FFSs and FRSs are different. In the case of FFSs, all plasma parameters (particle number density, thermal temperature, and velocity), along with the interplanetary magnetic field (IMF) increase. Conversely, in the case of FRSs, all solar wind and IMF parameters decrease, except the solar wind plasma velocity \citep{Landau1960,Burlaga1971a,Richter1985,Burlaga1995,Burguess1995,Tsurutani2011a}. See Figure 2 of \cite{Oliveira2017a} for comparisons between schematic profiles of FFSs and FRSs. These distinct shock dynamics are responsible for different magnetospheric responses to the impacts of FFSs and FRSs. \par
    
FFSs are known to be the most geoeffective class of IP shocks \citep{Echer2004a,Alves2011,Tsurutani2011a,Oliveira2015a, Oliveira2018a}. The first and most dramatic magnetospheric response to the impact of FFSs are characterized by positive sudden impulses (SI$^+$) in the horizontal component of the geomagnetic field measured by magnetometers located in the space \citep{Patel1970,Huttunen2005,Wang2010d,Su2015,Rudd2019} and on the ground \citep{Siscoe1968,Smith1986,Takeuchi2002b,Rudd2019}, as a consequence of a sudden increase of the solar wind dynamic pressure \citep{Russell1994a,Russell1994b,Echer2005a,Wang2010d,Rudd2019}. On the other hand, in the case of FRSs, the sudden impulse response is negative (SI$^-$) as geospace and ground magnetometers observe negative excursions in the horizontal component of the geomagnetic field \citep{Akasofu1964b,Takeuchi2002a,Andrioli2006,Zhang2010}. This is due to magnetospheric expansions caused by solar wind dynamic pressure decreases such as those associated with FRS/solar wind discontinuity impacts \citep{Joselyn1990,Araki1994,Zhang2010,Vichare2014}. \par  

 The impact of strong shocks can inject and energize particles into the radiation belts resulting in transient variation of the particle intensities. The classic example of such events is the strongest SI$^+$ recorded which occurred on 24 March 1991 \citep{araki1997anomalous}. The shock associated with this SI$^+$ formed a double-peaked inner radiation belt which lasted for years \citep{mullen1991double,looper1994observations}. The radiation belt electron and proton responses to shocks have been simulated by \cite{li1993simulation} and \cite{hudson1995simulation}. Their simulation results showed that electrons of few MeV were transported from L $> 6$ to L $\sim 2.5$ by an inductive electric field  which energizes them up to 40 MeV.  
A study by \cite{kanekal2016prompt} showed  that ultrarelativistic electrons with energies $> 6$ MeV were injected deep into the magnetosphere at L $\sim$ 3 within $\sim$ 2 min following the 17 March 2015 IP shock impact.  \cite{schiller2016prompt} carried out a statistical study of highly relativistic electron injections caused by IP shock interactions with the magnetosphere. They observed that not all their shocks show signatures in MeV electrons; around 25\% of IP shocks are associated with prompt  MeV electron energization, and 14\% are associated with prompt MeV electron depletions. Moreover, \cite{zhang2018observations} studied shock‐induced electric fields caused by IP shocks and suggested implications of these impulsive electric fields to the acceleration and transport of radiation belt electrons. Note that, the processes responsible for the enhanced/depleted intensities of energetic particles within the magnetosphere following shock impacts can be understood in terms of some combination of (i) enhanced particle source populations or their radial transport and (ii) localized acceleration. Although intense shocks are relatively rare, their impact on the inner radiation belt can last for years or even decades. Therefore, they can not be ignored in space weather and space technology related studies.

As can be seen, there has been considerable interest in studying the magnetospheric effects of SI$^+$s (caused by FFSs), but FRS interactions with radiation belt have received less attention. FRSs are generally not strong shocks however \cite{cattell2017dayside} showed that even weak FFS can disturbed the magnetosphere significantly, thus the study of FRS impact on magnetosphere is important.   Just as a compression of the magnetosphere due to FFSs results in various transient processes, a sudden rarefaction  due to FRSs or solar wind discontinuities can also generate interesting dynamics like generation of  transient currents in magnetosphere and ionosphere \citep{Vichare2014,shen2017dayside}. To the best of our knowledge, there are various studies reporting on the geomagnetic field response to SI$^-$; however, there is no comprehensive study of the radiation belt response to a FRSs. 

This gap in understanding motivates the present study to investigate an FRS impact on the outer radiation belt in detail. Although shock impact has been studied extensively using spacecraft observations and simulations, there are at least three important problems which still need attention: (i) Do reverse shocks affect the radiation belts significantly?, (ii) Do the impacts of reverse shocks on the inner and outer radiation belts simply mirror those of a forward shocks?, (iii) What are the primary drivers (or causes) responsible for transient variations in particle fluxes during the shocks? To address these questions, the present study carries out multi-point spacecraft observations to evaluate and compare the outer radiation belt responses to the rarefaction of the magnetosphere during the transit of a reverse shock across the Earth's magnetosphere. The observed variations in electron and proton fluxes are interpreted based on our current understanding of particle dynamics in the magnetosphere. 

Arrival of FRS at the Earth could make the radiation environment more harmful during and after shock transit. Intensification of high energy particle fluxes in radiation belt affect orbiting communication and navigation satellites adversely. If a high-intensity shock similar to the event of 24 March 1991 occurs in near future, its impact on space technology will be severe. The current study shows that even FRSs can affect radiation belt particles even though they are generally weaker than FFSs. The results from this study will assist in the better understanding of underlying physical processes and prediction of particle variations induced by FRS interactions with the Earth' magnetosphere. The study has further implications to shock-induced processes in other planetary magnetosphere of the solar system and exoplanets.
\section{Data and Methods}
We examined the interplanetary shock lists available at  (\url{http://ipshocks.fi/database}) and (\url{https://www.cfa.harvard.edu/shocks/ac_master_data/ac_master_2014.html}) to identify FRSs from 2013-2016, an interval when high quality continuous data are available from the Van Allen Probes and other spacecraft.

There were a total of 29 FRSs during this period. Here we present a detailed case study of the single event which strongest event on the list. This FRS was observed by the Wind spacecraft on 6 December 2014, at 21:09 UT, the spacecraft was located upstream in the solar wind at the time of the shock impact. To study the magnetosphere response GOES-13, GOES-15, LANL-97A, THEMIS, and Cluster spacecraft are used. 

The Wind Magnetic Field Investigation (MFI) instrument, consisting of two fluxgate magnetometers, was used for IMF observations \citep{lepping1995wind}. The Wind Solar Wind Experiment (SWE) Faraday Cup - Ion Data  was used to obtain information on solar wind plasma conditions \citep{Ogilvie1995}. We use the  MAGNETOMETER (MAG)  and MAGnetospheric Electron Detector (MAGED) instrument  on GOES-13 and 15  to understand the response at geosynchronous orbit  \citep{hanser2011eps,rodriguez2014intercalibration}.
MAGED is a set of nine collimated solid state detectors, data are provided as differential electron fluxes for  40, 75, 150, 275, and 475 keV midpoint energies from nine telescopes. The Synchronous Orbit Particle Analyser (SOPA) on the Los Alamos space environment monitor LANL-97A  in geosynchronous orbit ($\sim6.6 R_E$), is used to study electron and proton intensities in the nightside geosynchronous orbit. SOPA measures electrons in  nine electron channels from $\sim$ 0.05-1.5 MeV and protons from   $\sim$ 0.2-2.5 MeV. Note that neither LANL-97A nor any other LANL space environment monitor  carries any magnetometers \citep{taylor2004multisatellite}. The THEMIS Solid State Telescope (SST) records ions and electrons  within the energy range from 25 keV to 6 MeV. The  Flux Gate Magnetometer (FGM) on THEMIS measures the background magnetic field in the near-Earth space environment \citep{angelopoulos2009themis,auster2008themis}. The measurements of the IMF, solar wind parameters, and magnetospheric particle and field with one-minute cadence were utilized for the current study. 

\begin{figure}
	\centering
	\includegraphics[width=15.0cm]{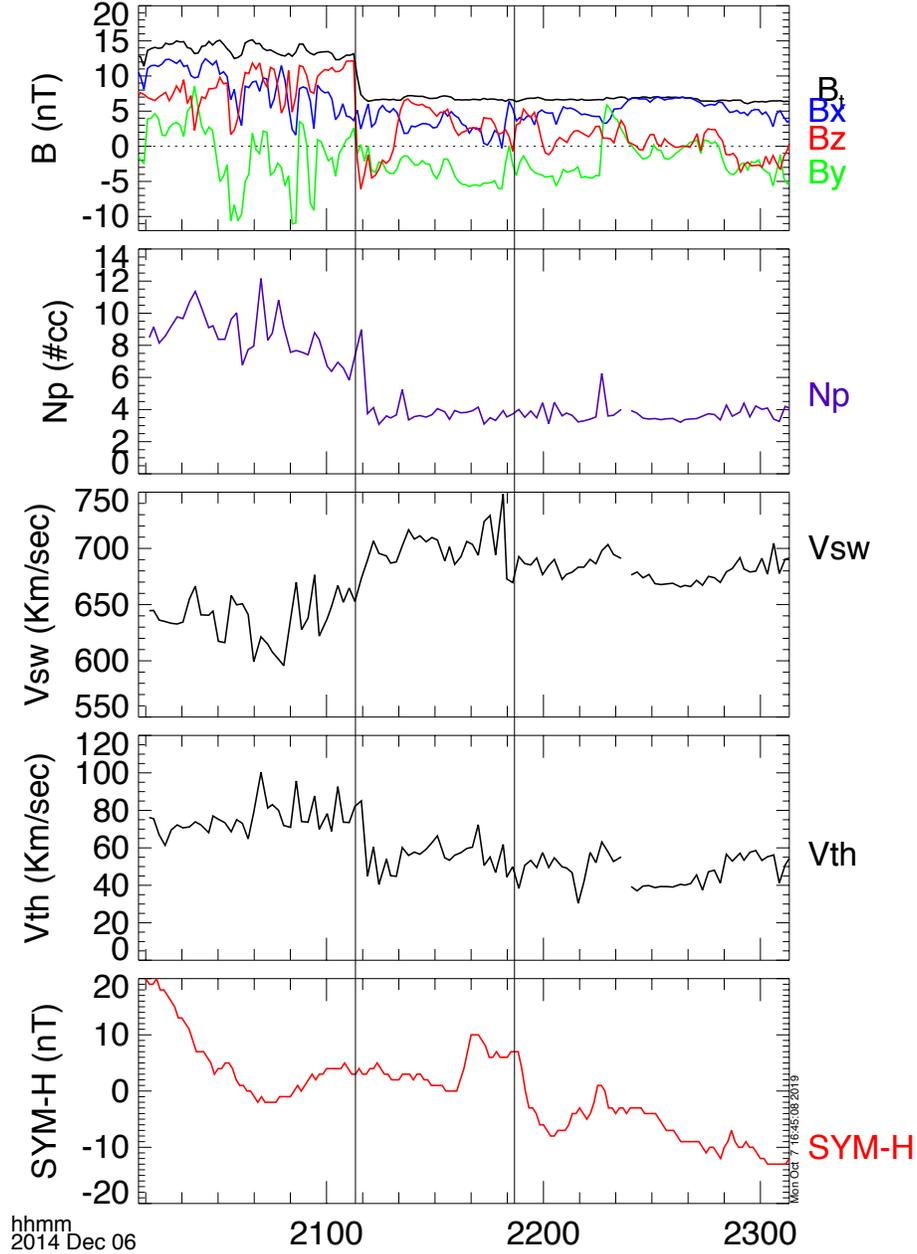}
	\caption{ Solar wind parameters observed by Wind during the FRS on 06 December 2014. The onset of the shock at Wind is shown by left vertical line and the right vertical line shows the  onset of the shock at the ground as seen in the SYM-H index. }
	\label{fig:IP_WIND}
\end{figure}

Figure ~\ref{fig:IP_WIND} shows IMF and solar wind parameters for the studied event. The top panel in the figure shows the total IMF strength (IMF, $B_t$) and its components (Bx, By, Bz). At the 21:09 UT shock onset, IMF Bz, thermal speed (Vth) and the solar wind particle density (Np)  decreased sharply, whereas the solar wind bulk (Vsw) shows the increase. These observed changes in the solar wind parameters clearly indicate the shock is a FRS. The Rakine-Hugoniot equations, based on assumptions of energy and momentum conservation across the shock front \citep{Landau1960,Tsurutani2011a,Oliveira2017a} yield the values 590 km/s and $\sim10^\circ$ for the shock speed and shock impact angle, respectively. Here, shock impact angle is the angle made by IP shock normal with the Sun-Earth line. The bottom panel of the figure shows SYM-H index, a measure of the global averaged; low latitude magnetic field response on the ground. The SYM-H index shows a decrease of about 20 nT around 21:54 UT which is consistent with the expected solar wind flow time $\sim45$ min from Wind's upstream location ($X_{GSE}=204.5 Re$, $Y_{GSE}= -89.5 Re$, $Z_{GSE}=-0.05Re$) to the Earth. 

\begin{figure}
\centering
\includegraphics[width=14.0cm]{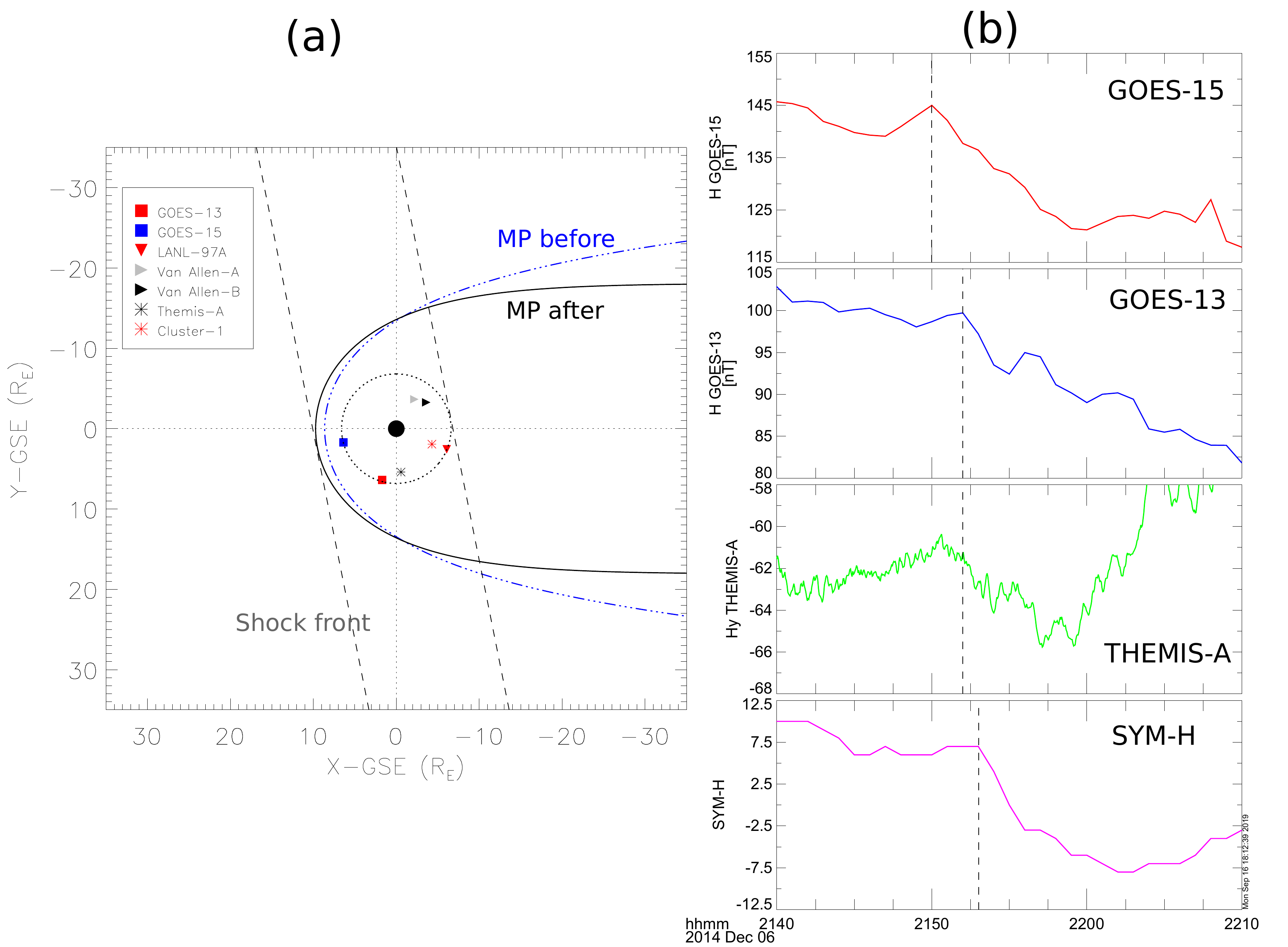}
\caption{(a)Spacecraft positions during the FRS event of 06 December 2014 at 21:50 UT. The magnetopause shapes just prior to (blue dashed line) shock arrival  and at the shock onset (black solid line) are shown. Dotted circle shows geosynchronous orbit. (b) Magnetic field variation caused by  the FRS event of 06 December 2014. the dashed vertical lines represent onset of the FRS as seen at the respective spacecraft. Total magnetic field of GOES is shown in top two panels. Third panel from top shows y component of magnetic field as recorder by THEMIS-A. Note that the y-component of the magnetic field (Hy) for THEMIS-A, is shown as other components did not show any clear variation. At the bottom  global low latitude average geomagnetic field response is shown by the SYM-H index.  }
\label{fig:sat_pos}
 \end{figure}
 
 Figure ~\ref{fig:sat_pos}a  shows magnetospheric spacecraft positions for the event. GOES-15 was located near noon, GOES-13 and THEMIS-A were located near dusk, and LANL-97A was near midnight. The figure also shows the magnetopause boundary estimated  by the \cite{shue1998magnetopause} magnetopause model just before (blue dashed curve) and just after (black solid curve) the shock impacted the magnetosphere. The shock wavefront is shown as inclined dashed black lines. The left and right lines indicate the location of the shock front at the time of magnetopause impact and at the time of the maximum depression observed in SYM-H associated with the shock i.e after 10  min of shock onset, respectively. Note that the magnetosphere expands in response to the FRS impact, as opposed to the compression caused by the impact of an FFS, as shown in figure 2 of  \cite{Rudd2019}.

\section{Observations and Results}
\subsection{Magnetic field response}

 
Figure ~\ref{fig:sat_pos}b shows the space-based and averaged global magnetic field response to this shock. The dashed vertical line indicates the onset of the FRS as seen in the magnetic field at GOES-15, GOES-13, THEMIS-A spacecraft and in SYM-H on the  ground. A clear decrease is seen at all spacecraft and on the ground.  The onset of the shock is seen earlier ($\sim$21:50:00) at GOES-15, later at GOES-13 ($\sim$ 21:52:00), and still later on the ground ($\sim$ 21:53:00). Timing analysis using GOES-15, GOES-13, and SYM-H index data indicates that the rarefaction front moved through the magnetosphere with a antisunward speed of $\sim$ $209-232km/s$. The observed speed is consistent with the reported speed of $\sim 300 km/s$ for an SI$^-$ by \citep{araki1988geomagnetic}.  THEMIS-A shows a gradual decrease in the magnetic field strength but the onset time is not very clear. After the onset of the shock observed at GOES-15 the magnetic field reaches a minimum in 10 minutes and after that almost remains steady. Unlike GOES-15, GOES-13 shows a steady decrease in magnetic field strength after the onset and the magnetic field continued to decrease. A clear local time dependence can be observed by comparison of the magnetic field profiles of GOES-15 and GOES-13.  Note that around 21:30 UT, prior to the FRS onset as seen at GOES-13,15 and SYM-H, there is a sharp enhancement in the magnetic field which could be due to a small density jump around 20:30 UT seen at Wind.

\subsection{GOES Particle flux response}

\begin{figure}
\centering
\includegraphics[width=15.0cm]{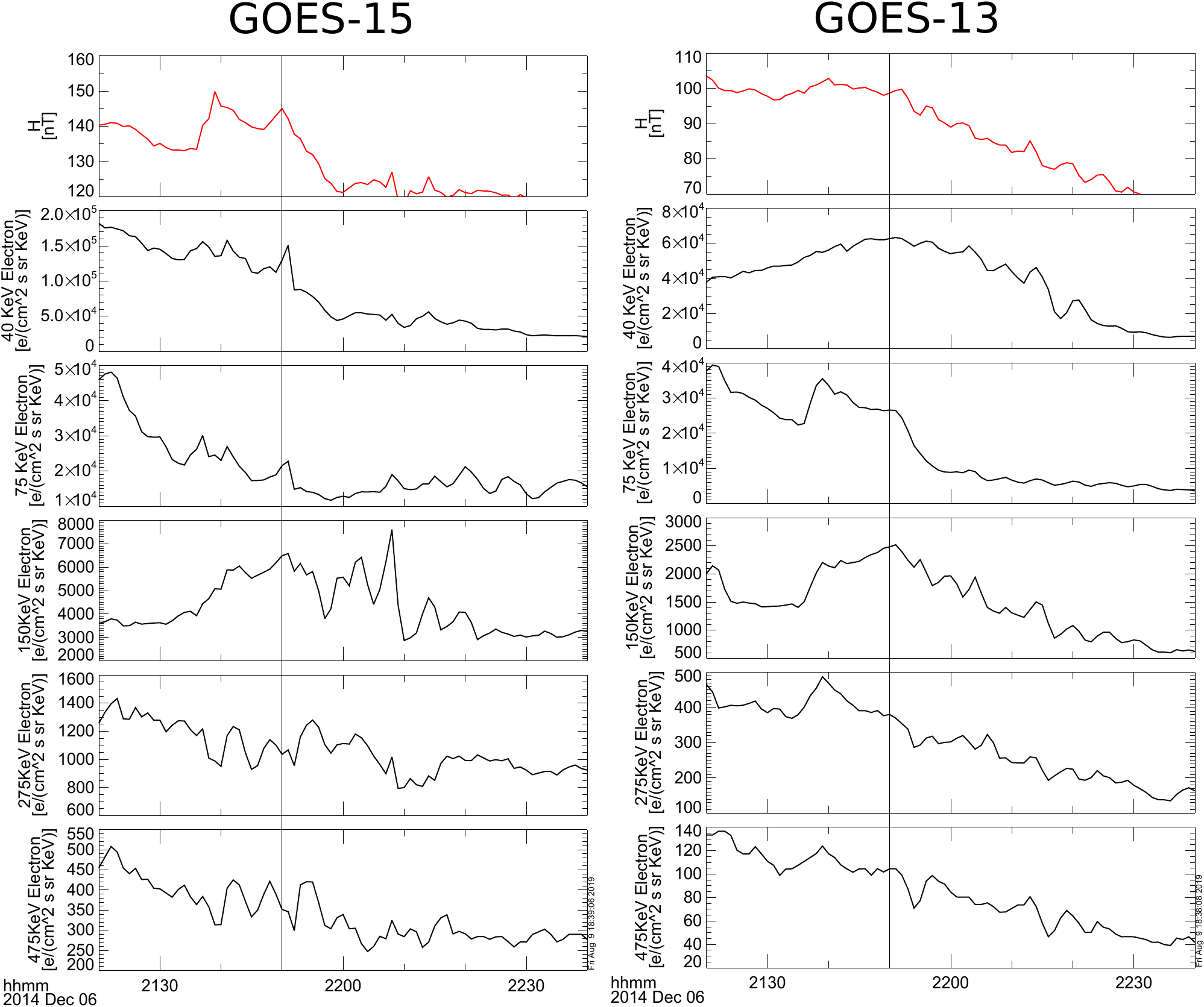}
\caption{Electron flux response to the FRS as observed by GOES-13,15. The vertical line indicate the onset of FRS signature as seen in magnetic field variations measured by GOES-15.  There is steady decreases in magnetic field at GOES-13 unlike GOES-15. The low energy particles show FRS signature whereas there is no clear signature a higher energies at both the spacecraft.}
 \label{fig:goesparti}
 \end{figure}
 
Figure \ref{fig:goesparti} shows GOES-15, 13 electron fluxes. The vertical line indicates the onset of the shock at each spacecraft as identified in Figure ~\ref{fig:sat_pos}. A decrease in electron flux for selected channels follows the rarefaction-induced magnetic field for the FRS duration. The following general characteristics can be readily noted based on observed signatures: (1) An almost instantaneous decrease is observed in most of the electron flux channels. (2) There is no indication of any dispersive effect over the entire spectral range as all the channels which show the signature show a decrease at the same time. (3) low energy (40-75  keV ) particle fluxes follow the magnetic field variations quite well.

However, close observation reveals strong energy dependent effects leading to significantly different signatures in the various energy channels. The lower energy channels show a rapid decrease as compared to the higher energy channels, which show weak or almost unidentifiable signatures. The signals at higher energies could be smeared out in time due to the fast drift rate superposing the time and spatial-variation.
Note that all electron fluxes show sharp changes around 21:30 UT prior to the FRS onset which could be associated with a small pressure jump around 20:30 UT observed by Wind. This is also evident in the GOES magnetic field and it is manifested as a  sharp increase in SYM-H around 21:35 UT.  Curiously, the flux profile seen in the 75 keV channel of GOES-13 is very well correlated with the magnetic field variation seen by GOES-15.  The percentage change in intensity for the different energy channels varies. For the 75 keV channel more than a 70\% decrease was recorded by GOES-13. The proton data were absent for both the spacecraft. 

\begin{figure}
\centering
\includegraphics[width=15.0cm]{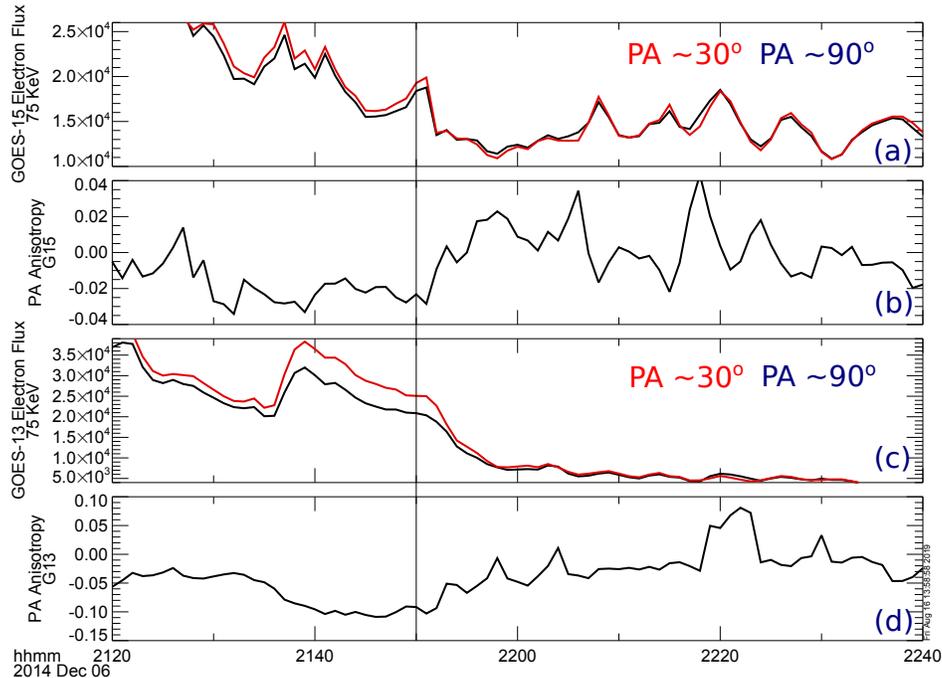}
\caption{Field-aligned (PA$\sim 30^o$) and perpendicular (PA$\sim 90^o$) pitch angle Electron flux response to the FRS as observed by GOES-13,15 for the 75 keV channel}
 \label{fig:goespitch}
 \end{figure}
 
Pitch angle distributions at various energies were examined. Intensities at all pitch angles decrease after the shock impact at both spacecraft. Figure \ref{fig:goespitch} shows near field-aligned ($\sim 30^o$) and perpendicular ($\sim 90^o$) electron fluxes of 75 keV electrons. Just prior to the shock arrival field-aligned intensities were greater than those for the equatorially trapped particles, in contrast to the nominal distribution. The anisotropy index (A) was calculated for both spacecraft, where $A= [j(90^o)-j(30^o)]/[j(90^o)+j(30^o)]$. The value of $A$ was negative prior to the onset of the FRS, and as an immediate response became less negative or even positive. After the shock $A$ remained close to zero, i.e almost isotropic distributions were observed. All other energies show a similar effect (not shown here). This implies that the particle population residing radially inward had an almost isotropic pitch angle distribution and that these particle populations moved to the spacecraft due to an expansion of the magnetosphere. Asymmetric or butterfly distributions are more commonly observed in the outer magnetosphere or higher L-shells than in the inner magnetosphere \citep{ni2016occurrence}.  Therefore, during the expansion of the magnetosphere caused by the FRS, a spacecraft could observe a decrease in anisotropy as the spacecraft encounters particles from the inner magnetosphere i.e as the distribution shifts from butterfly-like to one that is more isotropic.

\subsection{LANL Particle intensities response}
\begin{figure}
 	\centering
 	\includegraphics[width=16.0cm]{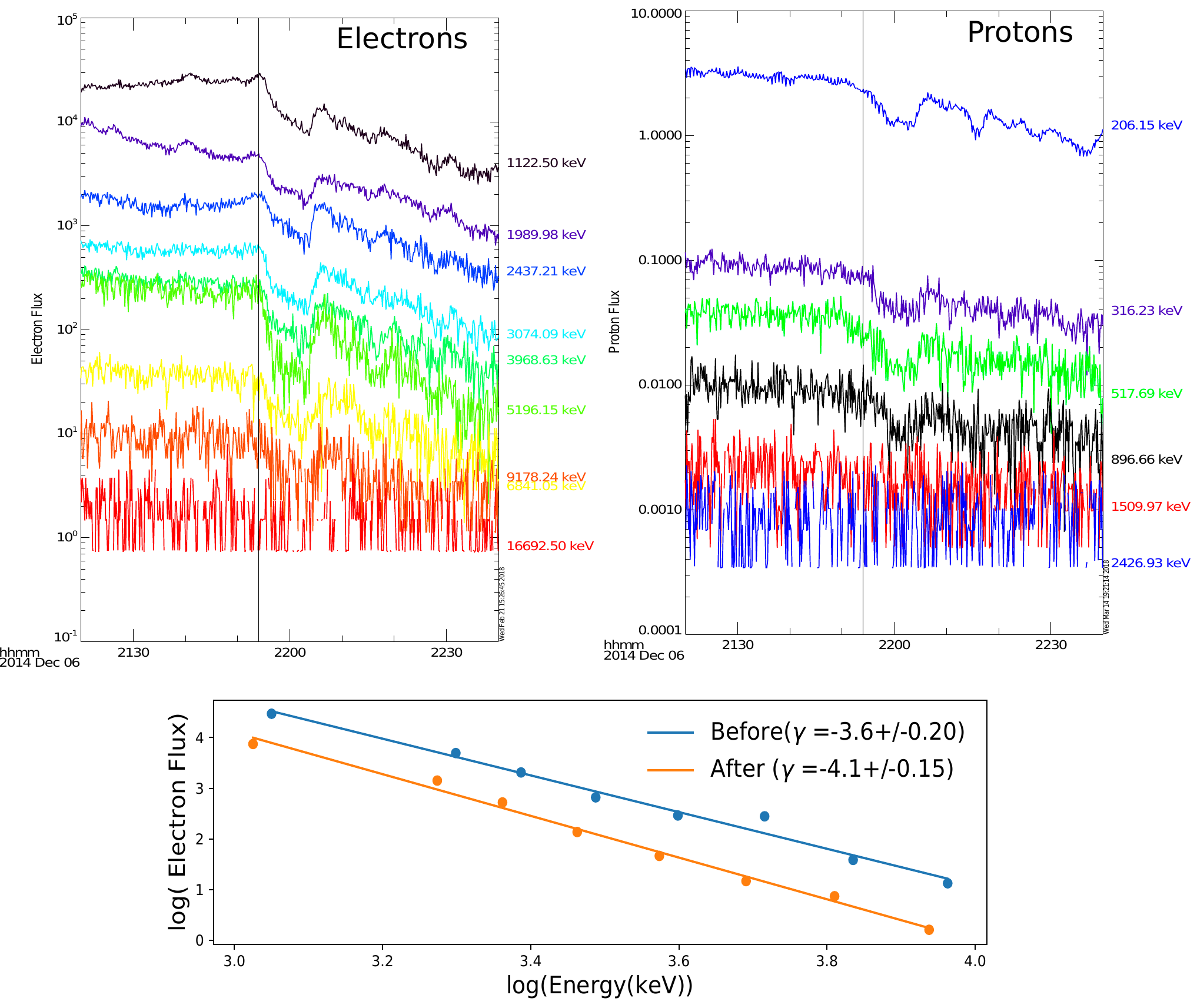}
 	\caption{Electron (Left panel) and Proton (right panel) flux  response to FRS as observed by LANL-97A. The vertical line indicates the onset of  the FRS signature as seen in particle intensities which coincide with the onset observed in the SYM-H index. Energy spectra of high energy electrons observed by LANL-97A (bottom panel) in night sector before and after the shock onset. }
 	\label{fig:electron}
 \end{figure}
 Figure \ref{fig:electron} shows the electron and proton fluxes observed by the LANL-97A spacecraft, located near midnight. A very strong and clear response is observed at this spacecraft in both the electron ($\sim$ 1-9 MeV) and proton ($\sim$0.2-0.9 MeV) fluxes. The observed decrease is non-dispersive as the onset is seen  simultaneously in all the energy channels. Interestingly, even MeV electrons show the  decrease in intensity at the onset time. The overall percentage of the decrease is $\sim 70\%$  for each electron channel and $\sim 47\%$ for the each proton channel which showed the signature.

 
As shown in Figure \ref{fig:electron} bottom panel, before and after electron spectra of electrons show that the spectral slope does change after the onset of the shock. The observed spectral slope for electrons is $\sim 3.6$ prior to the shock but has steepen to $\sim 4.1$ after the shock. The change in slope value is more than the error, indicating that the observed change is real and significant. The implications of this will be discussed in the Section 5.

\section{Model comparisons}
We use magnetospheric models to simulate the global response of the magnetosphere to the FRS and for comparison with the spacecraft observations. They can provide magnetic field responses at locations where no spacecraft observations exist.

The Block-Adaptive-Tree-Solarwind-Roe-Upwind-Scheme  BATS-R-US \citep{ridley2002model,toth2005space,toth2012adaptive} was used in the present study. This model was developed by the Computational Magnetohydrodynamics (MHD) Group at the University of Michigan and is now available as part of the Space Weather Modeling Framework (SWMF). The model run for the shock event was carried out at the Community Coordinated Modeling Center (CCMC). BATS-R-US has been used to investigate the interaction of IP shocks with the Earth's magnetosphere by many studies \citep{Koval2006,Ridley2006a,Samsonov2007,Samsonov2013,Samsonov2015}.

Solar wind plasma parameters (density, velocity, temperature) and IMF observed by the Wind spacecraft are inputs to the model. The solar wind data is transformed into GSM coordinates and propagated from Wind's position to the sunward boundary of the simulation domain. The Earth's magnetic field is approximated by a dipole and the dipole orientation updated during the simulation period.  Outputs of the run are magnetospheric plasma parameters including density, pressure, velocity, magnetic field, and electric currents.

 \begin{figure}[ht!]
 	\centering
 	\includegraphics[width=16.0cm]{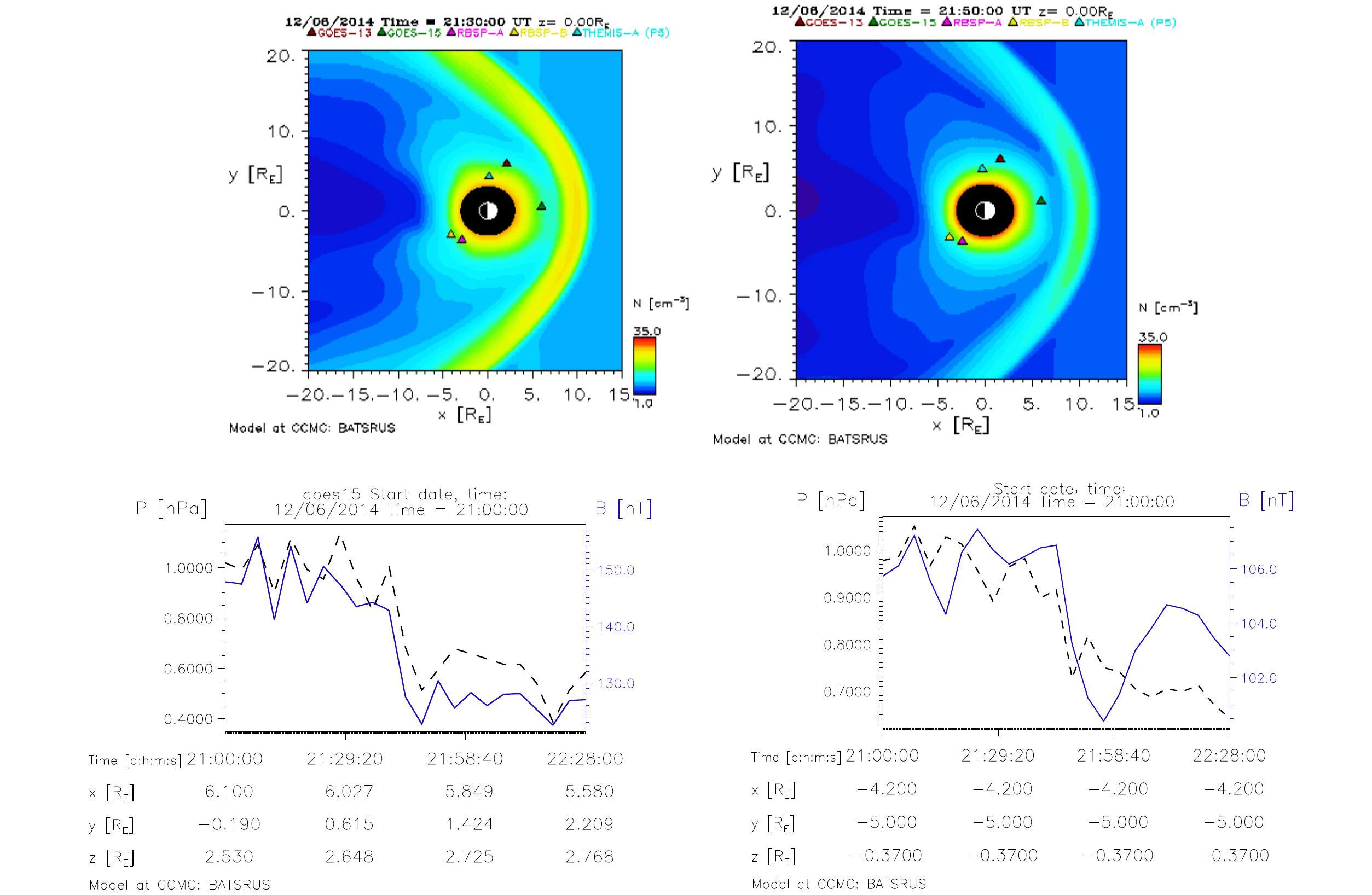}
 	\caption{BATS-R-US model outputs for the studied event. Top panels are magnetosphere density profiles in the equatorial plane. Bottom panels are time series outputs of pressure and magnetic field at GOES-15 and LANL-97A locations. }
 	\label{fig:model}
 \end{figure}
 
Figure \ref{fig:model} shows simulated densities in the equatorial plane of the magnetosphere before (left top panel) and after (right top panel) the shock arrival. The simulation reproduces the clear reduction in particle number density at geosynchronous orbit, an expected  FRS impact signature. The model time series for the magnetic field (B) and pressure (P) variations at the location of GOES-15 and LANL-97A are shown in the lower panels.  On the dayside the GOES-15 modeled-pressure shows a clear step like pressure drop, whereas the pressure decrease at LANL-97A is gradual. The model predicts  a step-like decrease of $\Delta B \sim 25 nT$ at GOES-15, with the magnetic field remaining low after the shock. Overall the model fields agree with GOES-15  observations to a reasonable degree. The modeled magnetic field at LANL-97A shows a decrease in $ B$ of just $6 nT$, very small compared to that at GOES-15.  There are no magnetic field measurements available from LANL-97A. So this model predication can not be validated using LANL-97A. Thus to validate the nightside model fields, we have comparedits predictions to the Cluster spacecraft observations (not shown here) which were taken in the nightside. 
The decrease of $\Delta B \sim 15 nT$ is observed by all four spacecraft. However, the BATS-R-US model predicts only $\sim 6$ nT variation at this location. This indicates that the model underestimates magnetic field values on the nightside. Therefore, it is likely that it also under-predicts magnetic field variations at LANL-97A and thus under estimates the adiabatic effect. The following section tries to quantify the contribution of adiabatic and other effects to the observed particle intensity variations.


\section{Contribution of adiabatic effects and radial gradients }
\begin{figure}[h!]
 	\centering
 	\includegraphics[width=16.0cm]{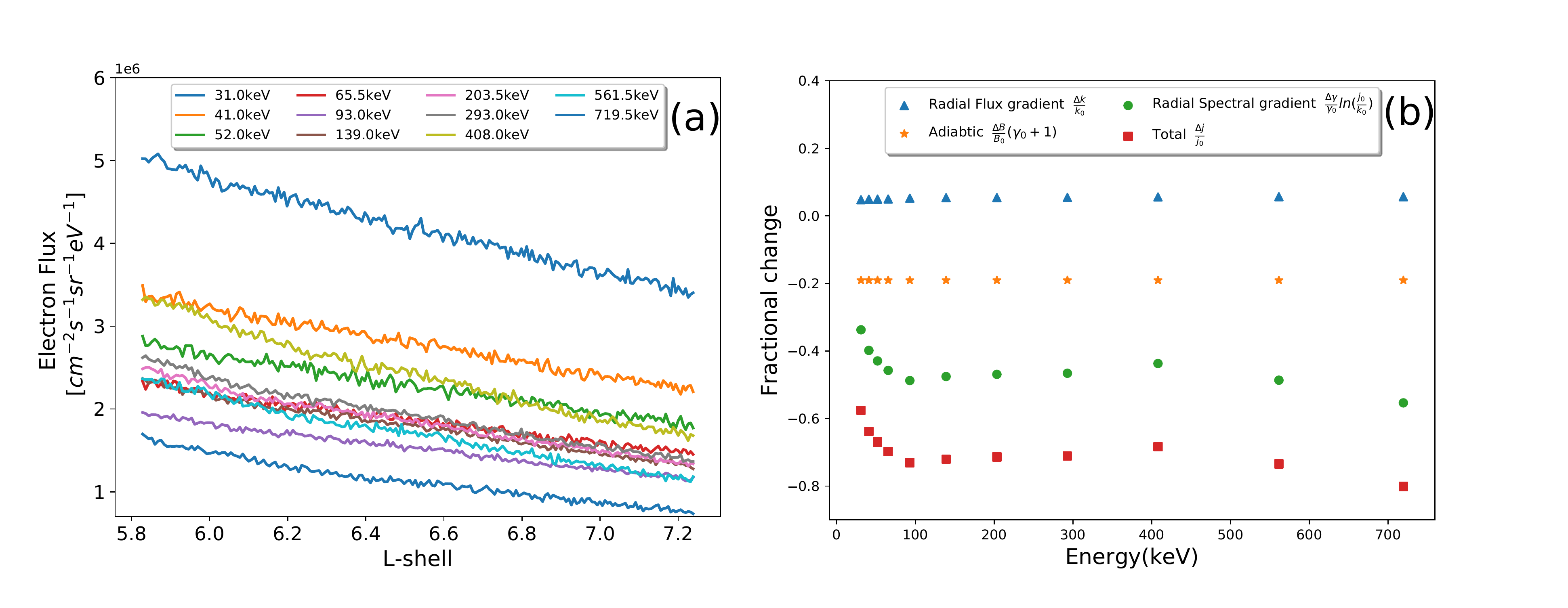}
 	\caption{(a)The L-shell variation of electron fluxes as observed by THEMIS-A during inbound 11:03 to 11:13 UT on 6th December 2014. (b) Estimated contribution of various terms associated with radial gradients and adiabatic effect in the observed decrease using THEMIS-A data. The red squares represent the net contribution of all three terms }
 	\label{fig:cal}
 \end{figure}

The variations in the particle intensities attending the shock  arrival could result from any combination of (1) advected radial gradients, (2) adiabatic effects,  (3) spectral changes, and other (4) non adiabatic processes.  From  \cite{wilken1986magnetospheric} the observed change in the flux (j) to first order estimate can be expressed by the following equation,  the first term represents the change due to adiabatic processes, whereas the second term represents the changes associated with advected radial gradients in the particle intensities.
\begin{equation}  \label{eq:2}
\Delta j (E=const)= \frac{\delta j}{\delta B} \Delta B +  \frac{\delta j}{\delta R} \Delta R\,,
\end{equation}
where $\Delta B$ and $\Delta R$ represent the changes in magnetic field strength and radial position seen by the magnetosphere plasma population.  If we assume the first invariant and phase space density are conserved during the transient rarefaction of the magnetosphere, this relationship can be expressed as 

\begin{equation}  \label{eq:3}
\frac{\Delta j}{j(E)} = \frac{\Delta B}{B_0} (\gamma_0 + 1)+ \frac{\Delta k}{k_0} +  \frac{\Delta \gamma}{\gamma_0} ln \Bigg(\frac{j_0}{k_0}\bigg)
\end{equation}
Here, the first term represents the change due to adiabatic processes, whereas the second term represents the changes associated with radial gradients and third due to change in spectral properties of the particle population. We have attempted to evaluate each term of this expression using THEMIS-A particle data. The orbit of THEMIS-A crossed a wide range of L-shells which enables estimation of gradients in particle  intensity and spectral index. These gradients are estimated around geosynchronous orbit prior to the onset of the shock. Figure ~\ref{fig:cal}a shows the electron flux data in nine energy channels utilized for the estimates.  Note that the higher fluxes are observed at lower L-shells  for these energies and locations. The flux measurements are used to calculate radial gradients in flux and spectral index. 

Thus, the various terms of Equation \ref{eq:3} are calculated based on these measurements. Since there are no magnetic field measurements onboard LANL-97A to estimate the adiabatic term, magnetic field observations by Cluster spacecraft $\Delta B =15 nT$ and $B_0=100nT$ were used. By using the TS04 model the magnetic foot points are estimated for pre-shock and post-shock solar wind dynamic pressure conditions to deduce a $\sim 0.5 R_E$ expansion of the magnetosphere at geosynchronous orbit, thus  $\Delta R$ was taken as $0.5 R_E$.

Figure ~\ref{fig:cal}b shows values for each term in Eq. \ref{eq:3} for several energies. The total fractional change ($ \frac{\Delta j}{j_0}$) is also presented to see the net estimate of the decrease in the various energy channels. The adiabatic ($\frac{\Delta B}{B_0} (\gamma_0 + 1)$) and spectral gradient ($\frac{\Delta \gamma}{\gamma_0} ln \big(\frac{j_0}{k_0}\big)$) terms are the dominant contribution to the observed decrease. Advection of the negative radial gradients ($\frac{\Delta k}{k_0}$) in intensity make only negligible positive contributions to the particle intensity.  The calculated total fractional decrease ($ \frac{\Delta j}{j_0}$)  for $100$ keV and above energies is $< -0.5$  reaches almost $-0.8$ for $700$ keV energy. This estimate is comparable to that observed. The observed fractional flux decrease at LANL-97 for $\sim 1122$ keV is  $-0.75$.  Whereas, at GOES-13 for 75 keV electrons the observed fractional decrease is $-0.71$ which is very close to the estimate of $-0.69$.  Note that these estimates are done using observed magnetic field, spectral slope and particle intensity gradients and remarkably we do get a close match with the observed fractional changes in the particle intensities. 

\section{Role of shock impact angle}
Van Allen Probes were located deep within the post midnight magnetosphere and did not see any significant variation in particle intensities or magnetic field. This could be due to a subdued response of magnetosphere to the shock at lower L shells.  Also, the magnetosphere might respond asymmetrically to an interplanetary shock depending on the angle of shock impact \citep{Oliveira2015a,Vichare2014,Oliveira2018a,Rudd2019}. The shock was slightly inclined, and first impacted the magnetosphere on the afternoon side which also could contribute to a weak or absent of shock impact as the Van Allen  Probes location . \par 

 \begin{figure}
 	\centering
 	\includegraphics[width=14.0cm]{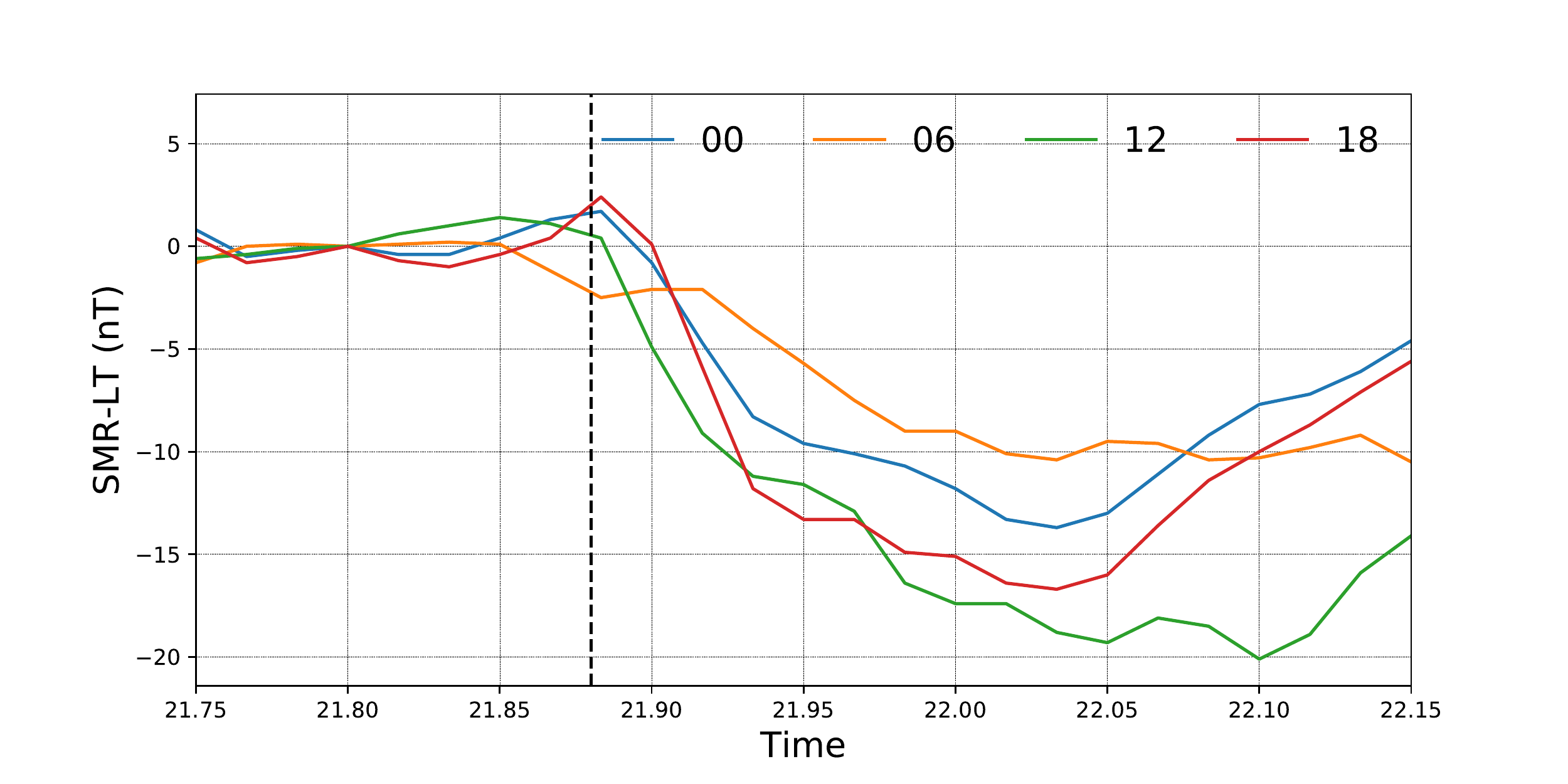}
 	\caption{Local time SMR indices during the event. The vertical dotted line represents the onset of  the FRS. whereas, the curves denote the local time SMR index at various local times. A strong local time dependence of the magnetic field decrease is associated with FRS.}
 	\label{fig:smr}
 \end{figure}
In order to infer the shock inclination influence we have used the Partial Ring current local time SMR index. SuperMAG partial ring current index (SMR) is similar to the SYM-H index but computed by a higher number of ground magnetometers \citep{newell2012supermag}.  Figure \ref{fig:smr} shows the various local time SMR indices. Four equally sized local time sectors are defined with centers at 00, 06, 12, and 18 MLT to compute the local time SMR which are shown in the figure. The maximum decrease is observed in the noon sector, whereas the minimum is seen in the dawn sector. The second strongest response is seen near the dusk-midnight sector. This implies that there is a strong local time dependence of the ground-based magnetic field response due to an asymmetric rarefaction of the magnetosphere. The Van Allen Probes were located in the dawnside  magnetosphere where the shock impact was weaker due to its inclination. Thus an asymmetric expansion of the magnetosphere could explain the observed  varied response of the particles to the shoversus local time .

\section{Discussion and conclusions}
The present study shows that not only  FFSs, but also FRS,  generate interesting dynamics. The geosynchronous response is investigated using GOES-13, 15 and LANL-97A spacecraft data. The magnetic field strength decreases at geosynchronous orbit and on the ground in response to the inclined FRS. The SYM-H index dropped by $\sim 20 nT$. The timing analysis shows that the rarefaction wave front propagated antisunward at $209-232km/s$. The particle response also seems to be global in nature as most electron energy channels from 10s of keV to MeV energies show a decrease in intensity. However, not all the energy channels show the same relative  decrease or variation in magnitude. It is possible that gradient and betatron effects counterbalance each other resulting in the absence or small size  of the signature or its absence in different energy channels which we have attempted to address here. The shock impact not only caused a decrease in particle intensities but also in pitch-angle anisotropy. This implies that those particle populations which advected outward through geosynchronous orbit due to the shock-induced magnetospheric expansion had an almost  isotropic pitch angle distribution. 

To identify the contribution of radial gradient and adiabatic effects to the observed decrease we followed the formalism  of \citep{wilken1986magnetospheric}. Moreover, the required magnetic field to carry out this calculation was inferred using the BATS-R-US model at the LANL-97 A location.  However, when we  validated the model outputs using the Cluster spacecraft magnetometer data we found that the model underestimated perturbations in the nightside magnetic field. Our first order estimate following \cite{wilken1986magnetospheric}, and using THEMIS-A  showed that the radial gradient makes only a small contribution. The work by \cite{west1973electron} and \cite{west1979signatures}  shows that the radial gradient of fluxes at constant energy tends to assume small values for low energy ($< 200 keV$) electrons and increase monotonically for higher energies. Our first order estimate following the \cite{wilken1986magnetospheric} approach  agrees with this, i.e the radial gradient values are very small ($\sim~0$) for keV electrons. Further, we find that the adiabatic and spectral terms significantly contribute to the observed decrease. The calculated contribution are consistent with the observed decrease in electron intensities at GOES-15,13 and LANL-97A spacecraft.

The shock impact on the magnetosphere was slightly inclined, resulting in an asymmetric expansion of the magnetosphere. This is well shown by variations in the SMR index at different local times. Spacecraft to spacecraft variations in the response at different local times could be explained by the inclined incidence of the shock. This highlights the importance of the shock impact angle during shock interaction with the magnetosphere and is consistent with the earlier reports, as shown by simulations \citep{Guo2005,Oliveira2014b,Samsonov2015} and observations \citep{Takeuchi2002b,Wang2006a,Oliveira2016a,Oliveira2018b}. \par

The adiabatic effect  underlying the observed decrease in relativistic electron intensities can be understood as follows.  The onset of the FRS causes the magnetosphere to expand, relativistic electrons conserving their third invariant move radially outward to compensate for the reduced local magnetic field  strength. Thus LANL-97 which is almost located at fixed radial distance then  measures the intensity of electrons previously present at lower L-shells. The outward moving electrons experience a weaker magnetic field and to conserve the first adiabatic invariant ($\mu$), their perpendicular energy must decrease. Also, the electrons will now see longer field lines which reduces their parallel energy.  Therefore, the distribution of electrons shifts to lower energy. A spacecraft measuring flux at fixed energy, $E_0$ and fixed radial distance measures the flux of electrons initially at a lower L-shell shifted to lower energies. Thus generally this outward movement results in a decrease of electron intensity at fixed energy for a decrease in magnetic field strength. In addition to this effect, if the  pre-shock magnetosphere had a strong gradient in spectral slope, even this also can contribute to the decrease. The dominance of the estimated spectral term supports this.  These two effects are primarily responsible for the observed decrease in particle fluxes caused  by the FRS on 06 December 2014.

Note that the magnetosphere responds differently to interplanetary FFSs and FRSs. The FFS compresses the magnetosphere,  whereas FRS causes an expansion of the magnetosphere. The present study shows that the reverse shock  predominantly caused the decrease in particle intensities. However, particle responses depend on magnetospheric conditions prior to the onset of the shock such as radial gradients in intensities and spectral index and could manifest as no effect or even enhancements during either fast forward or fast reverse shock. Even though FRS are rare, a statistical study would shed light on the general response of  magnetosphere particles to FRS. Moreover, the findings of the study have implications to understand the interaction of shocks with the other planetary magnetosphere.

\acknowledgments
The solar wind parameters, interplanetary magnetic field and geomagnetic indices used in this study are obtained from CDAWEB (\url{https://cdaweb.gsfc.nasa.gov/}) and SUPERMAG (\url{http://supermag.jhuapl.edu/}). We thank Wind, GOES, THEMIS, Cluster teams for providing the data. The modeling part of the work was carried out using the SWMF/BATSRUS tools developed at The University of Michigan Center for Space Environment Modeling (CSEM) and made available through the NASA Community Coordinated Modeling Center (CCMC). Also, we thank Community Coordinated Modeling Center (CCMC, Run Number \textnormal{\url{Ankush_Bhaskar_051118_1}}) for running the required models for this work. We thank Vassilis Angelopoulos, UCLA and 
Brian Kress, NOAA for valuable discussions.  This work is carried under NASA Living With a Star Jack Eddy Postdoctoral Fellowship Program, administered by UCAR’s Cooperative Programs for the Advancement of Earth System Science (CPAESS). A.B thanks UCAR, CPAESS and NASA/GSFC for providing excellent facilities and environment for the work. Some work at NASA/GSFC was funded by the Van Allen Probes mission.



\bibliography{DO_main.bib}
\bibliographystyle{aasjournal}



\end{document}